\documentclass[runningheads]{llncs}
\usepackage[T1]{fontenc}
\usepackage{graphicx,verbatim}
\usepackage{amssymb}
\usepackage{amsmath, amssymb, graphicx, booktabs}
\usepackage{multirow}
\begin{document}
\title{VIViT: Variable-Input Vision Transformer Framework for 3D MR Image Segmentation}

\titlerunning{VIViT for 3D MRI Segmentation}
%

\author{Badhan Kumar Das\inst{1,2}\thanks{Equal contribution} \and
Ajay Singh\inst{1,2 *} \and
Gengyan Zhao\inst{3 *} \and
Han Liu\inst{3} \and
Thmoas J. Re\inst{3} \and
Dorin Comaniciu\inst{3} \and
Eli Gibson\inst{3} \and
Andreas Maier\inst{2}
}
%
\authorrunning{Badhan Das et al.}
%
\institute{Siemens Healthineers, Erlangen, Germany \and
Friedrich-Alexander-Universität Erlangen-Nürnberg, Erlangen, Germany
\and
Siemens Healthineers, Princeton, New Jersey, United States\\
}


\maketitle              
\begin{abstract}


Self-supervised pretrain techniques have been widely used to improve the downstream tasks’ performance. However, real-world magnetic resonance (MR) studies usually consist of different sets of contrasts due to different acquisition protocols, which poses challenges for the current deep learning methods on large-scale pretrain and different downstream tasks with different input requirements, since these methods typically require a fixed set of input modalities/contrasts. To address this challenge, we propose variable-input ViT (VIViT), a transformer-based framework designed for self-supervised pretraining and segmentation finetuning for variable contrasts in each study. With this ability, our approach can maximize the data availability in pretrain, and can transfer the learned knowledge from pretrain to downstream tasks despite variations in input requirements. We validate our method on brain infarct and brain tumor segmentation, where our method outperforms current CNN and ViT-based models with a mean Dice score of 0.624 and 0.883 respectively. These results highlight the efficacy of our design for better adaptability and performance on tasks with real-world heterogeneous MR data. 


\end{abstract}

\section{Introduction}


Self-supervised pretrain techniques have been widely proven to be helpful in improving downstream tasks’ performance \cite{he2022masked,zhang2022dino,oquab2023dinov2}.
In medical imaging, accurately diagnosing a disease or segmenting a lesion often requires multiple modalities, especially for MRI, where multiple MR contrasts are usually needed. 
In real-world clinical MR data, different studies usually consist of different contrasts due to different acquisition protocols \cite{nael2021automated,hao2013multimodal}. 
This poses challenges for the current deep learning methods to do large-scale self-supervised pretrain, and to finetune on different downstream tasks with different input requirements, since these methods typically require all the input subjects/studies to have a fixed set of image modalities/contrasts. Hence, it is imperative to develop a flexible framework to handle variable set of modalities from each subject/study as input.


For medical image segmentation, Convolutional Neural Network (CNN) based models \cite{roth2018deep,malhotra2022retracted,ronneberger2015u,myronenko20193d} and Vision Transformer (ViT) \cite{dosovitskiy2020image,vaswani2017attention} based models \cite{manzari2023medvit,das2024co,dai2021transmed} have shown excellent performance. Models with a self-supervised pretrain phase, like UNETR \cite{hatamizadeh2022unetr,zhou2023self} and SwinUNETR \cite{hatamizadeh2021swin,tang2022self}, showed that self-supervised pretrain can improve the downstream task’s performance. However, by design these methods are not capable of handling datasets with variable input contrast sets during pretrain and finetune.

One related field is the missing modality in medical imaging \cite{azad2022medical}. However, recently a pioneer work AdaptiveUNETR \cite{das2025self} pointed out the difference between missing modality and variable input modality set. Typically, the missing modality methods need to know all the possible modalities during training, and the missing modality situations usually happen in the testing phase (training with the modality superset and testing on the subsets). Moreover, the missing modality models are not designed to handle unseen new modalities after training. In contrast, the variable input modality set issue needs to handle variable set of input modalities from each study during both training and testing, for both missing and previously unknown newly-added modalities, which is an underexplored field.

By extending the problem posed in AdaptiveUNETR, to fully leverage the power of self-supervised pretrain, a good system should have the following properties: \textbf{(1) all the available studies even having different sets of modalities/contrasts can be used in pretrain, so the data for pretrain can be maximized; (2) different downstream tasks with different input contrast requirements can all benefit from the same pretrained model.} These require the system being able to take variable set of modalities/contrasts from each study as input. 
In this work, we propose a variable-input ViT framework (VIViT) to fulfill this requirement. We summarize our contributions as follows:
\begin{itemize}
  

    \item We design a two-stage \textbf{dynamic patch tokenizer} that consists of a dynamic convolution for low-level feature extraction and another dynamic convolution to encode the extracted features to token vectors. It can adaptively encode different input contrasts with the same network structure.

   
    \item  We propose a \textbf{dynamic Transformer-Convolution encoder}. By leveraging the transformer’s ability of processing variable length of token sequence, the encoder is able to learn from variable input contrast sets. By integrating dynamic convolution layers into the transformer encoder, the local feature extraction is enhanced while the global attention mechanism is preserved.



    \item We introduce a two-step \textbf{modality fusion decoder}. First, the features are decoded within each modality with self-attention. Then the decoded features are aggregated across modalities and further decoded with a multi-level CNN decoder.


  \item With its novel architecture, the proposed VIViT framework flexibly handles variable input contrast sets. Extensive experiments show its abilities to learn from variable sets of contrasts in self-supervised pretraining and to transfer learned knowledge to finetuning with different input requirements, outperforming current methods in performance and adaptability.

\end{itemize}



\section{Methods}

\begin{figure}
\includegraphics[width=\textwidth]{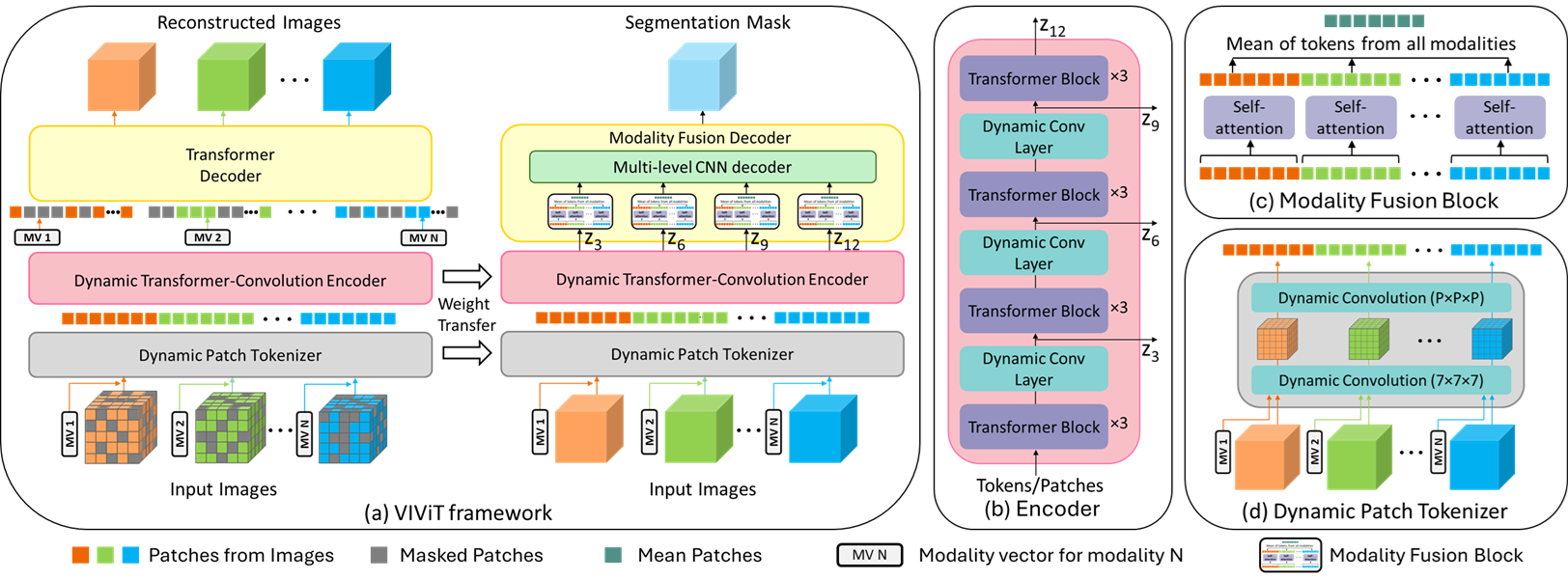}
\caption{Overview of the Variable-Input ViT (VIViT) framework. (a) The self-supervised pretrain (left) and downstream segmentation (right) architectures in VIViT framework (b) The dynamic Transformer-Convolution encoder (c) Modality fusion block, and (d) The dynamic patch tokenizer} \label{fig1}
\end{figure}

The overview of the proposed VIViT framework is shown in Figure \ref{fig1}. 
It contains two phases: self-supervised pretrain and downstream finetune. The architecture can effectively accommodate variable sets of input contrasts across different subjects/studies and between pretrain and finetune. For self-supervised pretrain, the masked autoencoder \cite{bao2021beit,he2022masked} strategy is used, where randomly masked image patches are reconstructed to learn the intra- and inter-contrast data distribution.
Then the pretrained network is finetuned for the downstream segmentation tasks.
We propose three key structures that enable the framework to handle variable sets of input contrasts both effectively and efficiently.

\subsection{Dynamic Patch Tokenizer}

We propose a dynamic patch tokenizer to convert input images of different contrasts to tokens with two-stage encoding (Figure \ref{fig1}-d). Both stages use dynamic convolution layers, which can dynamically adjust the kernel weights based on the input image type. 
For the first stage a large kernel size is used to extract low-level features \cite{luo2023lkd,lee20223d}, while in the second stage, the kernel size and stride are set equal to the patch size, enabling the conversion of the extracted features into 1D tokens.
Consequently, the same tokenizer can be used to tokenize different input contrasts with dynamically adjusted behavior.

 
Dynamic convolution \cite{liu2022moddrop++,das2025self} is utilized to facilitate contrast-specific adaptation of the tokenizer through the utilization of a modality vector. Each modality is assigned a unique vector \( \mathbf{m}_i \in \mathbb{R}^l \), projected linearly to generate weight \( \mathbf{w}_{conv} \) and bias \( \mathbf{b}_{conv} \) vectors, which are used to update the convolutional layer’s weights and biases, enabling modality-specific feature extraction \cite{yang2021unified,hu2022domain}. 
Given the original weights \( \mathbf{W} \) and biases \( \mathbf{B} \), the updated parameters are:

\[
\mathbf{W}_{updated} = \mathbf{W} \cdot \mathbf{w}_{conv}, \quad
\mathbf{B}_{updated} = \mathbf{B} \cdot \mathbf{b}_{conv}
\]

For a 3D input image of the ith contrast \( \mathbf{X}_i \in \mathbb{R}^{1 \times H \times W \times D} \) where $H,W,D$ are the height, width, and depth, respectively, convolution with updated parameters produces output:

\[
\mathbf{Y}_i = \text{Conv}(\mathbf{X}, \mathbf{W}_{updated}, \mathbf{B}_{updated})
\]


where \( \mathbf{Y}_i \in \mathbb{R}^{C \times \frac{H}{S}  \times  \frac{W}{S}  \times  \frac{D}{S} } \) represents the output of the dynamic conv layer. Here C is the output channel and S is the stride of the dynamic convolution.

Initially, for the first stage, a dynamic convolutional layer with $7 \times 7 \times 7$ kernel size and stride 2 is applied to extract features over a large receptive field from different input image modalities. This operation transforms each input contrasts of shape \( \mathbb{R}^{ 1 \times H \times W \times D} \) into a higher dimensional representation \( \mathbb{R}^{C \times \frac{H}{2} \times \frac{W}{2} \times \frac{D}{2}} \), where C denotes the number of the output channel of the dynamic convolution. 
Then, for the second stage, another dynamic convolution layer is used to generate tokens dynamically from the feature map of each input contrast. The patch size is used as the kernel and stride size, and the embedding dimension of the transformer is used as the output number of channels. The output is then flattened into 1D tokens and for each study all the tokens from all the modalities are used to form a long sequence of tokens before being sent to the transformer encoder.


\subsection{Dynamic Transformer-Convolution Encoder}

We propose a dynamic Transformer-Convolution encoder (Figure \ref{fig1}-b) to extract features from variable set of input contrasts for each study and learn the relationship of these contrasts. This design serves multiple objectives: 1) It takes advantage of the transformer’s ability of being able to process variable length of input sequence of tokens to handle variable set of input contrasts for each study. Meanwhile, the relationship of the contrasts can also be learned from the global self-attention. 2) By alternately utilizing transformers and dynamic convolution, the encoder can capture both local and global features. 3) After several transformer blocks, the 1D-token sequence of each modality will be reshaped to a 4D feature map (\( \mathbb{R}^{\text{T} \times \frac{H}{P} \times \frac{W}{P} \times \frac{D}{P}} \) where T is the embedding dimension of the transformer and P is the patch size) and processed by the dynamic convolution, where the spatial relationship of the features in the original 3D space are reinforced.


In our framework, the tokens of different contrasts from the same study are concatenated to form a long sequence and fed to the dynamic Transformer-Convolution encoder. A positional embedding and a modality embedding are added to each token based on the location of the 3D patch and the image modality respectively. The transformer block of the encoder applies a global self-attention across all the tokens from all the available modalities of each study to extract the global features and learn the inter-modality relationship. Dynamic convolution enables flexible local feature extraction from variable contrasts and takes the features' spatial relationship in the original 3D space into consideration. 
The encoder extracts features at multiple levels (Figure \ref{fig1}-b), and the multi-level feature maps (z3, z6, z9, z12) are later passed to the decoder through residual connections at the corresponding levels.



\subsection{Modality Fusion Decoder}



In the pretrain stage, a transformer decoder is used to decode the long sequence of feature tokens from each study’s all available modalities to reconstruct the masked input images \cite{he2022masked,das2025self}. In the finetune stage, to effectively decode the feature token sequence of variable length from variable sets of input contrasts, we propose a modality fusion decoder with a two-step approach. 


In the first step, the long sequence of encoded feature tokens from each study’s all available modalities are split into short sequences of each modality’s encoded feature tokens, and modality-specific self-attentions are used to decode these modality-specific short sequences. Then, the mean of the decoded the feature tokens is calculated across all the modalities to aggregate the decoded tokens at each position across modalities. This decoding step is done at all the encoding levels (z3, z6, z9 and z12). 
For each encoding level \( z_k \) (\( k \in \{3,6,9,12\} \)), the aggregated decoded token is computed as:

\[
F_j^{z_k} = \frac{1}{N} \sum_{i=1}^{N} \text{SA}_i^{z_k}(X_{i,j}^{z_k})
\]

Where \( X_{i,j}^{Z_k} \) is the \( j \)th (position) encoded feature token for modality \( i \) at level \( z_k \), \( \text{SA}_i^{z_k}(\cdot) \) denotes the self-attention layer for modality \( i \) at level \( z_k \), and \( N \) is the number of available modalities for each study.

In the second step, at each level \( z_k \), the aggregated decoded token \( F_j^{z;_k} \) at all the positions (\( j = 0, 1, \dots, J \)) are reshaped to a 4D tensor feature map ( \( \mathbb{R}^{\text{T} \times \frac{H}{P} \times \frac{W}{P} \times \frac{D}{P}} \) ), and all the 4D tensor feature maps at all the levels are further decoded and fused with a multi-level CNN decoder, like the UNET decoder \cite{ronneberger2015u,hatamizadeh2022unetr}.

This two-step approach is designed to improve the downstream segmentation tasks’ performance. It uses a modality-specific self-attention layer to adequately decode each modality’s encoded feature tokens at every level before fusing them and subsequently uses a multi-level CNN decoder to further decode and up-sample the feature map to the original input resolution to generate the segmentation mask.

\section{Experiments}

\subsection{Dataset}




\textbf{Brain Infarct Segmentation:} We utilized a dataset with 1,648 training, 193 validation, and 215 test studies, all annotated by expert radiologists for acute/
subacute brain infarct segmentation. 
All studies in this dataset include Trace-weighted (TraceW) and Apparent Diffusion Coefficient (ADC) images, while the T2-weighted (T2) image is an optional contrast that may or may not be present.
\textbf{Brain Tumor Segmentation:} We used the BraTS 2021 dataset \cite{baid2021rsna}, which consists of 1,251 brain MR studies. Each study includes four contrasts: Fluid Attenuated Inversion Recovery (FLAIR), native T1-weighted (T1), post-contrast T1-weighted (T1CE), and T2-weighted (T2) images, and each study is annotated with three tumor regions: tumor core (TC), whole tumor (WT), and enhancing tumor (ET). In our experiments, the dataset is split into 1,001 training, 100 validation, and 150 test studies.
\textbf{Self-supervised Pretrain:} An internal dataset with 45,374 MRI studies was used and each study has a variable number of MRI contrasts. 
For each study, all available contrasts from the following eight were used:
ADC, TraceW, T2, Gradient Echo (GRE), Susceptibility-Weighted Imaging (SWI), T1, T1CE, and FLAIR.

\subsection{Implementation}


The proposed VIViT framework was implemented with MONAI \cite{cardoso2022monai} and PyTorch frameworks. All experiments were done on 4 NVIDIA A100 GPUs (40GB). For self-supervised learning (SSL), a weighted Adam optimizer, an L2 loss function, and a cosine annealing scheduler with an initial learning rate of 1e-5 were used. The SSL pretraining was done for 500 epochs with a 70\% masking ratio and a batch size of 4. The downstream brain infarct and tumor segmentation tasks were trained with a Dice loss \cite{milletari2016v}, a weighted Adam optimizer, and a cosine annealing scheduler with an initial learning rate of 1e-4 for 200 and 300 epochs, respectively. 
3D input volumes were normalized to have zero mean and unit standard deviation based on non-zero voxels and were resized to 128 × 128 × 128.
The best validation weights were used for testing, and the Dice similarity coefficient was used for evaluation.

\subsection{Results}

For segmentation tasks, we compared the performance of VIViT with UNET \cite{ronneberger2015u}, UNETR \cite{hatamizadeh2022unetr}, SegResNet \cite{myronenko20193d}, SwinUNETR \cite{hatamizadeh2021swin}, and AdaptiveUNETR \cite{das2025self} on brain infarct and tumor segmentation. For the models not designed to handle variable sets of input contrasts, a zero-filled tensor is used when any modality is unavailable in a study (e.g. in infarct segmentation, T2 may be absent for some studies). As shown in Table 1, our proposed method outperformed all the compared methods and achieved the best mean Dice score (0.624 for brain infarct segmentation and 0.883 for brain tumor segmentation).

\begin{table}
\caption{Performance comparison of different methods on brain infarct segmentation (mean Dice score) and brain tumor segmentation (TC, WT, ET and mean Dice score). }
\label{tab:segmentation}
\centering
\begin{tabular}{c c c c c c c c}
\hline
Model  & Infarct Segmentation & \multicolumn{4}{c}{Tumor Segmentation} \\
\cline{3-6}
& Mean Dice & Dice TC & Dice WT & Dice ET & Mean Dice \\
\hline
UNET  &  0.370 & 0.819 &	\textbf{0.909} &	0.806 & 0.845 \\
UNETR  & 0.576   & 0.838 & 0.850 & 0.858 & 0.849 \\
SegResNet & 0.587 & 0.854 & 0.875 &	0.835 & 0.855 \\
SwinUNETR  & 0.611  & \textbf{0.873} & 0.883 & 0.867 & 0.874 \\
AdaptiveUNETR & 0.561 & 0.797 & 0.871 & 0.831 & 0.833 \\
AdaptiveUNETR+SSL & 0.598 & 0.814 & 0.869 & 0.833 & 0.839 \\
VIViT  & 0.596  & 0.862 & 0.892 & 0.882 & 0.879 \\
VIViT+SSL  & \textbf{0.624}  & 0.871 & 0.893 & \textbf{0.884} & \textbf{0.883}  \\
\hline
\end{tabular}
\end{table}



Notably, VIViT with self-supervised pretraining improves the segmentation performance over VIViT without pretraining  by 2.8\% on brain infarct and by 0.4\% on brain tumor, demonstrating the effectiveness of pretraining in enhancing downstream performance. 
Our VIViT framework can learn from studies with different sets of contrasts during pretraining, which gives it the ability to leverage as many studies as available (> 45,000 studies in our experiment) to significantly benefit downstream tasks. Furthermore, the pretrained model is highly effective in transferring knowledge across different segmentation tasks, despite variations in input requirements and contrast availability.


\begin{table}
\caption{Model performance for different input contrast set situations between pretrain and finetune (the downstream task’s input contrast set is a subset (SSL) or a joint set (SSL Partial) of pretrain’s input contrast set).}
\label{tab:addition}
\centering
\begin{tabular}{c c c c c c c c}
\hline
Model & Infarct Segmentation & \multicolumn{4}{c}{Tumor Segmentation} \\
\cline{3-6}
& Mean Dice & Dice TC & Dice WT & Dice ET & Mean Dice \\
\hline
VIViT & 0.596  & 0.862 & 0.892 & 0.882 & 0.879 \\
VIViT+SSL partial & 0.615   &  0.876 & 0.894 & 0.886 & \textbf{0.885} \\
VIViT+SSL & \textbf{0.624}  & 0.871 & 0.893 & 0.884 & 0.883 \\
\hline
\end{tabular}
\end{table}

To further study the impact of VIViT’s pretrain when the downstream task has new contrasts that are unseen in pretrain (pretrain and finetune’s input contrast sets are joint sets), we conducted another self-supervised pretrain (SSL partial in Table 2), excluding all available T1 and T2 contrasts from the pretrain dataset. The pretrained model is finetuned and tested on the same downstream tasks’ data (where infarct has T2 and BraTS has T1 and T2 as unseen contrasts in pretrain). The results are shown in Table 2. Despite the unseen input contrasts in pretrain, the pretrain can still benefit the downstream tasks by transferring the knowledge learned from their common contrasts (intersection of the joint sets), while incremental knowledge about the new contrasts only seen in finetune and their relationship with the common contrasts can be “picked up” during finetune. Also, Table \ref{tab:addition} shows that for brain infarct segmentation the benefit of pretrain is greater when it covers all the downstream task’s input contrasts (SSL in Table 2, finetune’s input contrast set is a subset of pretrain’s).


\subsection{Ablation Studies}






Table \ref{ablation_mask} presents an ablation study assessing the impact of the proposed structural components as their progressive integration into our framework. The baseline VIViT model with only the modality fusion decoder achieves a mean Dice score of 0.595 on brain infarct segmentation. After replacing the original ViT tokenizer \cite{dosovitskiy2020image,zhou2023self} with the proposed two-stage dynamic patch tokenizer, the performance improves to a mean Dice score of 0.611, indicating the effectiveness of the 2-stage tokenization and the benefit of feature extraction with a large convolution kernel. Moreover, replacing the Transformer encoder with the proposed dynamic Transformer-Convolution encoder further improves the performance to a mean Dice score of 0.624, demonstrating the new encoder's superior performance by extracting local and global features with the spatial relationship of the features taken into consideration.
In addition, we investigated the effect of different masking ratios in self-supervised pretrain on the downstream task. Table \ref{ablation_mask} shows that with 70\% masking ratio, VIViT achieved the best mean Dice score on brain infarct segmentation. A too-high or too-low masking ratio may negatively affect the downstream task’s performance.

\begin{table}
\caption{Ablation study on structural components and masking ratio used in self-supervised pretraining on brain infarct segmentation.}\label{ablation_mask}
\centering
\begin{tabular}{ccc}
\hline
Category & Configuration & Mean Dice Score\\
\hline
\multirow{3}{*}{Model Components} & VIViT (Modality Fusion Decoder)  & 0.595 \\
& + Dynamic Patch Tokenizer  & 0.611 \\
& + Dynamic CNN-Transformer Encoder  & \textbf{0.624} \\
\hline
\multirow{3}{*}{Masking Ratio} & 50\%  & 0.602  \\
& 70\%  & \textbf{0.624} \\
& 90\%  & 0.613 \\
\hline
\end{tabular}
\end{table}

\section{Conclusion}

In this work, we propose the VIViT pretrain and finetune framework to effectively improve deep learning model’s adaptability and performance on the real-world heterogeneous MR data and tasks. The newly designed dynamic patch tokenizer, dynamic Transformer-Convolution encoder and modality fusion decoder enable the proposed framework to efficiently process variable set of input contrasts in each study during pretraining and finetuning. Through extensive experiments on brain infarct and brain tumor segmentation, we demonstrate that our approach’s superior performance over the current CNN and ViT based methods. Overall, the results highlight the potential of the proposed framework on handling variable input in medical imaging. Future work may include investigating both supervised and self-supervised pretrain strategies on the proposed framework and extending its application to other tasks in medical imaging.

\clearpage
\bibliographystyle{splncs04}
\bibliography{mybibliography}
%





\end{document}